\documentclass[review]{elsarticle}

\usepackage{lineno,hyperref}
\usepackage{color}
\usepackage{amsmath,amsfonts,amssymb,esvect,ulem}
\usepackage{float}
\usepackage{graphicx}
\usepackage{epsfig}
\usepackage{cancel}
\usepackage{epstopdf}
\usepackage{subcaption}
\modulolinenumbers[5]

\journal{Journal of Computational Physics}

\long\def\symbolfootnote[#1]#2{
	\begingroup\def\thefootnote{\fnsymbol{footnote}}
	\footnote[#1]{#2}\endgroup}

\newcommand\blankfootnote[1]{%
	\let\svthefootnote\thefootnote
	\textheight 1in
	\let\thefootnote\relax\footnotetext{#1}%
	\let\thefootnote\svthefootnote%
}

\usepackage[stable]{footmisc}
\newcommand{\sfootnote}[1]{\footnote{#1}}
\usepackage{manyfoot}
\newfootnote{B}
\makeatletter
\def\sfootnote#1{\ifx\protect\@typeset@protect
	\FootnotetextB{}{#1}%
	\else
	\relax
	\fi
}

\newcommand{\bea}{\begin{eqnarray}}
\newcommand{\eea}{\end{eqnarray}}
\newcommand{\be}{\begin{equation}}
\newcommand{\ee}{\end{equation}}
\newcommand{\beas}{\begin{eqnarray*}}
\newcommand{\eeas}{\end{eqnarray*}}
\newcommand{\bdm}{\begin{displaymath}}
\newcommand{\edm}{\end{displaymath}}
\newcommand{\bl}{\bss\begin{itemize}}
\newcommand{\el}{\vspace{-.5\baselineskip}\end{itemize}\ess}
\newcommand{\benu}{\bss\begin{enumerate}}
\newcommand{\eenu}{\vspace{-.5\baselineskip}\end{enumerate}\ess}
\newcommand{\bfg}{\begin{figure}}
\newcommand{\efg}{\end{figure}}
\newcommand{\bt}{\begin{table}}
\newcommand{\et}{\end{table}}
\newcommand{\bc}{\begin{center}}
\newcommand{\ec}{\end{center}}
\newcommand{\btb}{\begin{center}\begin{tabular}}
\newcommand{\etb}{\end{tabular}\end{center}}
\newcommand{\bss}{\singlespacing}
\newcommand{\ess}{\doublespacing}
\def\arbspace#1{\def\baselinestretch{#1}\@normalsize}

\newcommand{\bas}{\begin{arbspace}}
\newcommand{\eas}{\end{arbspace}}
\newcommand{\bi}{\begin{itemize}}
\newcommand{\ei}{\end{itemize}}
\newcommand{\ben}{\begin{enumerate}}
\newcommand{\een}{\end{enumerate}}

\newcommand{\lop}{\mathcal{L}}
\newcommand{\mv}{\mathcal{V}}
\newcommand{\mj}{{\Gamma}}
\newcommand{\lopv}{\lop v}
\newcommand{\lopp}{\lop\psi}

\newcommand{\vr}{\vec{r}}

\makeatletter

\newcommand{\Rmnum}[1]{\expandafter\@slowromancap\romannumeral #1@}
\makeatother

{\par\noindent%
 \rule[0pt]{\linewidth}{0.2pt}
 \vspace*{-9pt}
 \linespread{0.0}\small\verbatim}%
{\rule[-5pt]{\linewidth}{0.2pt}\endverbatim}

{\par\noindent%
 \rule[0pt]{\linewidth}{0.2pt}
 \vspace*{-9pt}
 \linespread{1.0}\scriptsize\verbatim}%
{\rule[-5pt]{\linewidth}{0.2pt}\endverbatim}

\newcommand{\st}{\sigma_\mathrm{t}}

\DeclareMathAlphabet\mathbfscr{OMS}{cmsy}{b}{n}

\newcommand{\psii}{\psi^\mathrm{inc}}
\newcommand{\psiinc}[1]{\psi^\mathrm{inc}_{{#1}}}

\newcommand{\omen}{\ome\cdot\nabla}

\newcommand{\e}[1]{\ensuremath{\times 10^{#1}}}
\newcommand{\TAMU}{Texas A\&M University}

\newcommand{\sigt}{\sigma_\mathrm{t}}
\newcommand{\sigs}{\sigma_\mathrm{s}}

\newcommand{\dome}{d\Omega}

\newcommand{\intdin}{\int\limits_{\ndo<0}\dome\ \int\limits_{\pd}ds\ }

\newcommand{\tm}{$\times$}

\newcommand{\sss}[1]{S$_{#1}$}

\newcommand{\fp}{4\pi}

\newcommand{\pd}{{\partial\mathcal{D}}}

\newcommand{\intli}[1]{\int\limits_{{#1}}}

\newcommand{\ndo}{\vec{n}\cdot\ome}
\newcommand{\absndo}{\left|\ndo\right|}

\newcommand{\lo}{L$_1$}
\newcommand{\lt}{L$_2$}

\newcommand{\intfpd}{\int\limits_{4\pi}\dome\ \int\limits_\mathcal{D}d\vec{r}\ }

\newcommand{\qs}{q_\mathrm{s}}

\newcommand{\ome}{\vec{\Omega}}



\begin{document}

\begin{frontmatter}

\title{Effectively Non-oscillatory Regularized L$_1$\ Finite Elements for Particle Transport Simulations}

\author[mymainaddress]{Weixiong Zheng\corref{mycorrespondingauthor}}
\cortext[mycorrespondingauthor]{Corresponding author}
\ead{zwxne2010@gmail.com}

\author[mymainaddress]{Ryan G. McClarren}
\ead{rgm@tamu.edu}

\address[mymainaddress]{Nuclear Engineering, \TAMU,~College Station, TX 77843-3133}

\begin{abstract}
	In this work, we present a novel regularized \lo\ (R\lo)\ finite element spatial discretization scheme for radiation transport problems. We review the recently developed least-squares finite element method in nuclear applications and then derive an \lo\ finite element by minimizing the \lo\ norm of the transport residual. To ensure stability, we develop a consistent \lo\ boundary condition (BC). Our method requires a nonlinear solve that we treat as a series of weighted least-squares calculations. The numerical tests demonstrate that our method removes the numerical artifacts that arise in LS solutions in problems with voids and pure absorbers. In these problems, our method gives non-oscillatory solutions. We also show that only a few iterations of the nonlinear solver give significant improvement over the standard LS solution.
\end{abstract}


\end{frontmatter}


\section{Introduction}
{
	Neutral particle transport problems are governed by the  linear Boltzmann transport equation, a first-order, hyperbolic equation for the phase space density of particles. The streaming operator in the transport equation is a linear advection operator. Consequently, problems where particles travel long distances between scattering interactions, so-called streaming-dominated problems, e.g. void and strong absorber problems, can have discontinuous solutions, a fact that spatial discretization schemes should respect. On the other hand, in regions where particles travel very short distances between scattering interactions, the transport equation asymptotically limits to a  diffusion equation. There has been much research into methods that are asymptotic preserving for this limit \cite{larsen1987asymptotic, larsen1989asymptotic}. At present the most widely used spatial discretization scheme that can handle both discontinuous solutions in the streaming dominated case and preserves the asymptotic diffusion limit  is the discontinuous finite element method (DFEM)\cite{adams_asym,mcc_pnasymp}. DFEM is widely used despite the notorious disadvantage of requiring additional degrees of freedom compared with the continuous finite element method (CFEM) that is commonly used in elliptic and parabolic problems.
	
Another approach to solve transport problems recasts the transport equation into a symmetric, second-order form. This can be done by deriving even-parity equations and self-adjoint angular flux equations\cite{vincent_dissertation,zheng_dissertation,lewis1993, hanuvs2016use,morel_saaf} or by using the least-squares finite element method. The least-squares finite element method seeks solutions in a finite element space to minimize the squared residual of the transport equation. The minimizer of the squared residual is the solution to the weak form of a symmetrized transport equation. Thereafter, CFEM can be used to solve this transport equation. Recent work has successfully applied least-squares finite elements to a variety of transport applications\cite{yaqi_invite,zheng_invite,laboure:2016,zheng_cdls,Hansen:2015jq,morel_holo,yaqi_physor16,clifmc,ketelsen2015least}.
	
There remain a number of open problems regarding least-squares finite elements. For instance, using an under-resolved mesh in problems with strong absorbers can induce oscillations and negative particle densities\cite{zheng_l1sn,zheng_l1pn,pete_nonnegative,petermc}. These negative densities, in addition to being non-physical, yield nonsensical reaction rates, which are usually the primary quantity of interest for particle transport simulations. Moreover, in multi-D situations, void or near-void situations can also induce negativity in the particle density due to oscillations caused by the existence of discontinuities in the solution to the continuum equations. These oscillations exist even in solutions to the first-order transport with DFEM\cite{petermc} when piecewise linear or high-degree basis functions are used. Even without oscillations, void regions can induce low accuracy of the least-squares method without specific corrections \cite{laboure:2016}. Yet, in real-world problems, like radiation shielding problems or remote sensing, those situations are usual and inevitable.
	
	For the least-squares (LS) method, one cause of the oscillations is that the L$_2$\ norm of transport residual overestimates the contribution from large residual components. Like least-squares fitting in data analysis, when trying to fit every single data point, the fitting principle would overweight the contribution from large-error points, leading to erroneous and oscillatory results\cite{zheng_mc15,bnjiang_l1}. One of the remedies is to develop finite elements that minimize the residual of the transport equation in other norms, such as \lo. This is difficult because the \lo\ norm is non-smooth and typically requires the solution of nonlinear equations. Previously, Jiang developed the iteratively reweighted least-squares (IRLS) method for linear advection\cite{bnjiang_l1}.\ The IRLS method uses a power of the inverse residual as the weighting function for a least-squares method. For simplicity, IRLS uses an approximation to the residual in the weighting function and assumes the weighting function to be constant in each spatial cell. The rationale behind this choice is that such a method would approximate the \lo\ norm-induced method. It was shown that for discontinuous boundary data, IRLS  is extremely accurate in the interior of the domain. However, Lowrie and Roe \cite{lowrie_l1}\ demonstrated that IRLS does not necessarily propagate information correctly from boundary so that if the incident boundary condition (BC) is smooth, IRLS can give erroneous results. In particular,  large gradients of the solution can be mistakenly treated as discontinuities. Furthermore, Lowrie and Roe demonstrated that IRLS is not an \lo\ method. 
	
	In a more recent work, by developing efficient nonlinear solving techniques, Guermond approximated the \lo\ solution for several problems in fluid dynamics based on Newton's method\cite{guermond_l1}.\ The \lo\ method is demonstrated to be accurate and stable in problems where least-squares has difficulty and oscillations and accurately treats smooth solutions in contrast to Lowrie's findings about IRLS. Yet, Guermond's implementation of \lo\ does not have a continuum form, making it difficult to apply to problems of particle transport where strong interactions could dominate advection. It also lacks a theory for consistent boundary conditions, which is required for many transport problems. 

We wish to have a method with stability property of the \lo\ method, as demonstrated by Guermond, with an implementation that builds on the technology recently developed for least-squares finite elements. Therefore, we introduce a regularized \lo\ method for solving particle transport problems. Such a method is designed to be a regularized version of \lo\ method in the sense that \lo\ method will be used only when the pointwise residual becomes larger than certain criteria otherwise the least-squares method is used. 
Additionally, the scheme is designed to be compatible with source iteration and acceleration techniques like diffusion synthetic acceleration (DSA)\cite{alcouffe_dsa} that are de rigeur for efficiently solving transport problem. 
We also develop a consistent \lo\ BC which, as we will demonstrate, is necessary to obtain accurate solutions.
	
	The remainder of this paper is organized as follows: in Sec.\ \ref{s:ls0},\ we derive the LS weak formulation solving one-speed steady-state radiation transport equation; we then  derive an \lo\ and further a regularized \lo\ (R\lo)\ finite element method in a continuum form in Sec.\ \ref{s:l1}.\ Therein, we further derive the consistent \lo\ and R\lo\ boundary weak formulation to prevent oscillations on incident boundaries. We briefly discuss the details of the implementation. Thereafter, we demonstrate the method efficacy in Sec.\ \ref{s:numerics}.\ We conclude the work in Sec.\ \ref{s:conc}.
}

\section{Least-Squares Method for One-speed Neutron Transport Equation}\label{s:ls0}
{
	\subsection{One-speed transport equation}
	{
		Given that we are interested in spatial discretization of the transport equation, we will restrict ourselves in this work to steady state, energy-independent transport problems, with isotropic scattering and volumetric fixed source. Extending our method with time and energy dependence and anisotropic scattering is relatively straightforward using existing methods.
		
		The steady, mono-energetic transport equation is given by \cite{adams_iron_water,glasstone,zheng_dissertation}:
		\begin{align}\label{eq:transport}
		\omen\psi(\vr,\ome)+\st(\vr)\psi(\vr,\ome)=\frac{\sigs\phi(\vr)}{\fp}+\frac{{Q}(\vr)}{4\pi}, \qquad \vr \in \mathcal{D}
		\end{align}
		where $\ome \in \mathbb{S}_2$ is the directional vector on the unit sphere corresponding to the direction of travel for particles; $\psi$\ is the angular flux with units of particles per unit area per steradian per second; $\sigs$ and $\st$ are respectively the scattering and total cross sections with units of inverse length; $\phi$\ is the scalar flux defined as \[\phi(\vr)=\intli{\fp}\dome\ \psi(\vr,\ome);\] $Q$\ is the isotropic volumetric fixed source.

The boundary conditions for Eq.~\eqref{eq:transport} specify $\psi$ for incoming directions on the boundary:
\[ \psi(\vr, \ome) = \psii (\vr,\ome), \qquad \vr \in \partial \mathcal{D},\quad \ome \cdot \hat{n} < 0,\]
where $\hat{n}(\vr)$ is the unit outward normal on the boundary of the domain.
		
		For notational simplicity, we will also use the operator form as the following:
		\begin{align}
		\lop\psi=\qs,
		\end{align}
		where $\lop$\ is the streaming plus removal operator \[ \lop \equiv \omen+\st,\] and the total (fixed plus scattering) source is \[\qs\equiv \frac{\sigs\phi(\vr)}{\fp}+\frac{{Q}(\vr)}{4\pi}.\] 
		
	We will discretize the angular component of the transport equation using the discrete ordinates (S$_N$) method \cite{lewis1993} where we use a quadrature set for the angular space, $\{w_n, \ome_n\}$, to obtain $N$ equations of the form
	\[ \lop_n \psi_n = \qs,\]
	where $\psi_n = \psi(\vr, \ome_n),$
	 \[ \lop_n \equiv \ome_n \cdot \nabla +\st,\] 
	 and 
	 \[ \phi(\vr) \approx \sum_{n=1}^N w_n \psi_n(\vr).\]
	}
	On the boundary we write $\psi_n(\vr) = \psii(\vr, \ome_n)$ for $\vr \in \partial\mathcal{D}$ and $\ome_n \cdot \hat{n} < 0$. 
	
	This choice of angular discretization will not affect the derivation of our spatial discretization. Therefore, we will drop the $n$ subscripts in the following sections. For example, we could use our regularized \lo\ method with the spherical harmonics treatment of the angular variable. 
	
	\subsection{Interior Weak Form of the Least-Squares Discretization}\label{s:ls}
	{ 
		To derive a LS discretization of the transport equation, we begin with defining the \lt\ norm of transport residual $R=\lop\psi-q_\mathrm{s}$ away from the boundary of the domain:
		\begin{align}\label{e:ls_funcin}
		\mj_{\mathrm{L}_2}(\psi)=\intfpd R(\psi)^2,
		\end{align}
		where we have restricted $\psi$\ to belong to a finite element space $\mv$.
		
		We next introduce an arbitrarily small perturbation $\epsilon v$,\ where $\epsilon >0 $\ is an arbitrarily small number and $v$\ is a weight function in the finite element space $\mv$.\ Then we obtain the perturbed functional:
		\begin{align}
		\mj_{\mathrm{L}_2}(\psi+\epsilon v)=\intfpd\left(R(\psi+\epsilon v)\right)^2.
		\end{align}
		To minimize the functional, we expect the first derivative to be zero in order to find the stationary point in the finite element space \cite{runchang},\ i.e.
		\begin{align}
		\frac{\partial\mj_{\mathrm{L}_2}}{\partial\epsilon}\bigg|_{\epsilon=0}=\intfpd\lopv\left(\lopp-\qs\right)=0.
		\end{align}
		Therefore, after rearranging, we have the weak formulation for the interior: find $\psi \in \mathcal{V}$ such that for any $v \in \mv$
		\begin{align}\label{e:lsw0}
		\intfpd\lopv\lopp=\intfpd\lopv\qs.
		\end{align}
	}
	
	\subsection{Boundary condition and complete weak formulation}
	{
		It is straightforward as well to obtain a weak form for the BC for the LS method. Similar to Eq.\ \eqref{e:ls_funcin},\ we define the functional for the BC measured by the L$_2$\ norm:
		\begin{align}
		\mj_{\mathrm{b,L}_2}=\intdin\lambda\absndo\left(\psi-\psiinc{}\right)^2,
		\end{align}
		where $\lambda$\ is a cross section related multiplier and defined as:
		\begin{equation}\label{e:choice}
		\lambda=\begin{cases}
		\sigt,&\sigt>0,\\
		1.0,&\mathrm{otherwise}.
		\end{cases}
		\end{equation}
		For the non-void situation, the LS weak form is globally conservative\cite{morel_holo,zheng_dissertation}\ with the choice in Eq.\ \eqref{e:choice}.\ The first choice of $\lambda$ in void is somewhat arbitrary, but has been observed to be adequate.	
		With the same procedure as in Sec.\ \ref{s:ls},\ we arrive at the boundary weak form:
		\begin{align}
		\intdin\lambda\absndo v\psi=\intdin\lambda\absndo v\psiinc{}.
		\end{align}
		Combining this result with Eq.\ \eqref{e:lsw0},\ we reach the complete LS weak form:
		\begin{align}
		\intfpd\lopv\lopp+\intdin\lambda\absndo v\psi=\nonumber\\
		\intfpd\lopv\qs+\intdin\lambda\absndo v\psiinc{}.
		\end{align}
	}
}

\section{Derivation in $L_1$\ Norm}\label{s:l1}
{
	\subsection{Smoothed \lo\ norm\ and \lo\ finite element method}\label{s:lo_int}
	{
		Similar to the LS method, we begin by defining the \lo\ norm of the transport residual:
		\begin{align}
		\mj_{\mathrm{L}_1}(\psi)=\intfpd|R(\psi)|
		\end{align}
		A suitable finite element method would be developed by minimizing the functional above. With the procedure  introduced in Sec.\ \ref{s:ls},\ we obtain a perturbed functional:
		\begin{align}\label{e:l1_func}
		\mj_{\mathrm{L}_1}(\psi+\epsilon v)=\intfpd|R(\psi+\epsilon v)|,\ v\in\mv,\ \epsilon>0
		\end{align}
		
		However, the functional $\mj_{\mathrm{L}_1}$ is not differentiable at the points where $R=0$. 
		To find the fixed point of Eq.~\eqref{e:l1_func} we develop an approximate \lo\ norm, as in \cite{zheng_l1pn,zheng_l1sn}.		
		For a small number $\zeta$ we approximate
		\begin{equation}\label{e:l1}
		|R|\approx\sqrt{R^2+\zeta^2}.
		\end{equation}
		Figure\ \ref{f:l1-smoothed}\ illustrates the effect of the approximation for different values of $\zeta$. With decreasing $\zeta$,\ $\sqrt{x^2+\zeta^2}$\ converges to $|x|$\ rapidly. 
		\begin{figure}[ht!]
			\begin{center}
				\includegraphics[width=.7\textwidth]{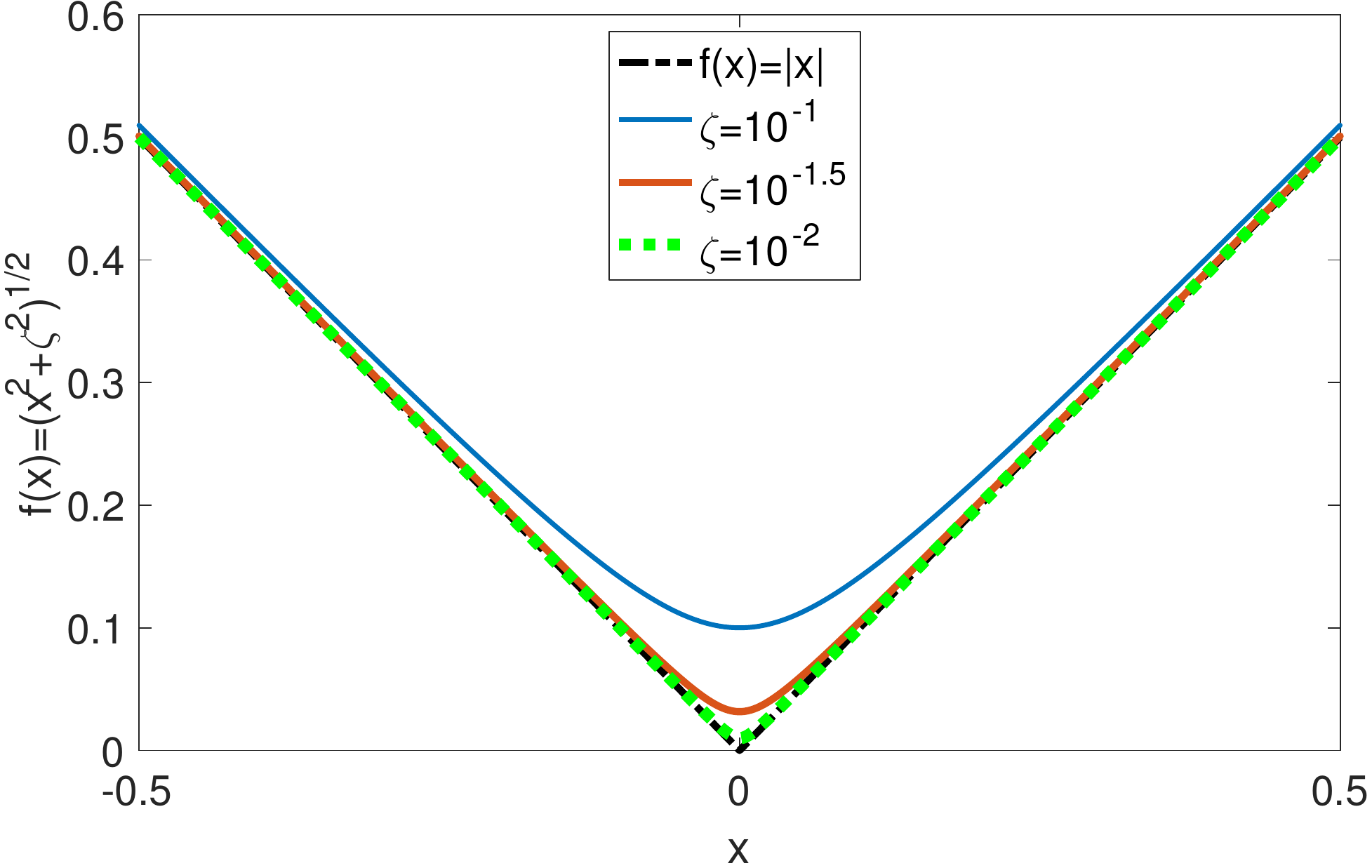}
				\caption{$\sqrt{x^2+\zeta^2}$\ vs $|x|$\ for different $\zeta$\ values.}
				\label{f:l1-smoothed}
			\end{center}
		\end{figure}
		
		Introducing Eq.\ \eqref{e:l1}\ into \eqref{e:l1_func},\ we obtain a differentiable approximation to the  perturbed \lo\ norm functional:
		\begin{align}\label{e:l1_smooth}
		\mj_{\mathrm{L}_1}(\psi+\epsilon v)\approx\intfpd\sqrt{R(\psi+\epsilon v)^2+\zeta^2}.
		\end{align}
		To minimize the convex functional, in the finite element space $\mv$,\ we need to find its stationary point where the derivative is zero:
		\begin{align}
		\frac{\partial\mj_{\mathrm{L}_1}}{\partial\epsilon}\bigg|_{\epsilon=0}=\intfpd\frac{\lopv(\lopp-q_\mathrm{s})}{\sqrt{R^2+\zeta^2}}=0.
		\end{align}
		Taking the limit $\zeta\to0$,\ the smoothed \lo\ expression limits to the \lo\ weak form as $\sqrt{R^2+\zeta^2}\to|R|$ for $|R| \neq 0$:
		\begin{align}
		\intfpd\frac{\lopv(\lopp-q_\mathrm{s})}{|R|}=0,
		\end{align}
		or equivalently,
		\begin{align}
		\intfpd\frac{\lopv\lopp}{|R|}=\intfpd\frac{\lopv q_\mathrm{s}}{|R|}.
		\end{align}
		This is equivalent to the weak form for the LS finite element with a weighting function that depends on the solution. Clearly, this weak formulation is nonlinear because the evaluation of $|R|$ requires the solution $|\psi|$. 
		
	}
	
	\subsection{An \lo\ BC}
	{	
		The na\"{i}ve BC for the \lo\ method would use BC from the LS weak form. However, we have observed that the application of this BC causes stability problems on the incident boundaries. A hypothesis is that the norms measuring residuals on the boundary and the interior should be consistent.
		
		We can derive a regularized \lo\ BC similar to the approach used for the interior functional. Namely, we write
		\begin{align}
		\mj_{b,\mathrm{L}_1}=\intdin\lambda\absndo\left|\psi-\psiinc{}\right|.
		\end{align}
		A boundary weak form as the following can be achieved through similar minimization process as above:
		\begin{align}
		\intdin\lambda\absndo\frac{v\psi}{\left|\psi-\psiinc{}\right|}=\intdin\lambda\absndo\frac{v\psiinc{}}{\left|\psi-\psiinc{}\right|}.
		\end{align}
	}
	
	\subsection{\lo\ and regularized \lo\ weak forms}
	{
		Whence we have the complete \lo\ finite element weak formulations:
		\begin{align}
		\intfpd\frac{\lopv\lopp}{|R|}+\intdin\lambda\absndo\frac{v\psi}{\left|\psi-\psiinc{}\right|}\nonumber\\
		=\intfpd\frac{\lopv q_\mathrm{s}}{|R|}+\intdin\lambda\absndo\frac{v\psiinc{}}{\left|\psi-\psiinc{}\right|}
		\end{align}
		
		Due to the assumption of letting $\zeta$\ vanish, the weak form above is an exact \lo\ finite element formulation. However, solving such a weak form can be extremely challenging, especially when residual in the problem varies by several orders of magnitude and when the residual vanishes in certain regions. Instead, we propose a regularized \lo\ formulation using a factor $\theta>0$:
		\begin{align}\label{e:rl1}
		\intfpd\frac{\theta \lopv\lopp}{\max(\theta,|R|)}+\intdin\lambda\absndo\frac{\theta v\psi}{\max(\theta,\left|\psi-\psiinc{}\right|)}\nonumber\\
		=\intfpd\frac{\theta \lopv q_\mathrm{s}}{\max(\theta,|R|)}+\intdin\lambda\absndo\frac{\theta v\psiinc{}}{\max(\theta,\left|\psi-\psiinc{}\right|)}.
		\end{align}
		This regularization is such that in regions with moderate to large residuals, i.e.\ $|R|>\theta$, the regularization factor \[\frac{\theta}{\max(\theta,|R|)} \rightarrow \frac{\theta}{|R|}\]  and the \lo\ weak form is used. On the other hand, when the residual is small, the regularization factor goes to 1 and the least-squares weak form is used.
		
		Numerical experiments indicate that the solution is largely insensitive to the choice of $\theta$.\ The effects of varying it will be illustrated in Sec.\ \ref{s:void}.\ Normally, we choose $\theta$\ to be around $0.01|R|_\mathrm{max}$, where $|R|_\mathrm{max}$\ denotes the maximum absolute residual of all direction and space. Smaller $\theta$\ can be used, yet, we observe the efficiency of the linear solver is degraded without a concomitant gain in solution accuracy. 
		
		In our implementation the residuals are evaluated on each spatial quadrature point. This is one of the drawbacks of this method in that it requires storing or performing on-the-fly calculations of the point wise residuals. 
	}

    \subsection{Nonlinear solution method}
    {
    	As R\lo\ is nonlinear, an appropriate nonlinear scheme is necessary in order to solve Eq.\ \eqref{e:rl1}. 
 We will use a Picard iteration scheme that evaluates the residual using the estimate of $\psi$ from the previous iteration. The scheme is initialized using the unweighted least-squares solution. The steps in the solution are
    	\begin{enumerate}
    		\item[1.] Calculate pointwise residuals for Nonlinear Iteration (NI) $l$\ from NI $l-1$;
    		\item[2.] Update the weak form Eq.\ \eqref{e:rl1};
    		\item[3.] Solve Eq.\ \eqref{e:rl1}\;
    		\item[4.] Given a nonlinear tolerance $tol$,\ check nonlinear convergence $e=\frac{\|\phi^l-\phi^{l-1}\|}{\|\phi^l\|}$:
    		\begin{enumerate}
    			\item[(1)] If $e<tol$,\ stop.
    			\item[(2)] else, go to Step 1.
    		\end{enumerate}
    	\end{enumerate}
    	In the solution of Eq.\eqref{e:rl1} source iteration with diffusion synthetic acceleration (DSA) is utilized\cite{lewis1993,Hansen:2015jq}. Though R\lo,\ as well as LS, does not have consistent low order diffusion acceleration scheme\cite{Hansen:2015jq},\ we have found that DSA is still effective. Developing a consistent DSA scheme for LS and R\lo\ should be the target of future work. 
    }
}

\section{Numerical Results}\label{s:numerics}
{
	All numerical results below used finite element solutions  carried out with the C{\tt ++}\ Open source library {\tt deal.II}\cite{dealii84}.
	In this section, we present four 2D test problems {with bilinear finite elements on rectangular meshes}. We first investigate the behavior of R\lo\ in void with discontinuous incident BC.

		\subsection{Void problem}\label{s:void}	
As mentioned above transport problems in voids can contain numerical artifacts such as oscillations and negative solutions due to the discontinuous nature of the analytic solution \cite{pete_nonnegative,zheng_l1sn,zheng_l1pn,petermc}. A demonstration of these phenomena can be seen in the solution to a beam problem at a grazing angle entering a void from the boundary. The solution to this problem is discontinuous with solution being zero outside the volume within the view of the beam. Schemes such as LS will have Gibbs oscillations due to the discontinuity. Figure\ \ref{f:void-pcolor}\ shows the LS and R\lo\ solution for this a 2-D void with where the domain is a 0.5$\times$0.5\ cm square. Along part of the boundary from $x=0.1$ to $0.3$ cm there is a unit incident angular flux at the angle $\ome=(1/\sqrt{3},1/\sqrt{3},1/\sqrt{3})$.\ The unit incident BC is applied only on part of the boundary. The LS solution oscillates near the discontinuity, and has an overshoot of 7\%. The R\lo\, solution, on the other hand, is monotone and non-negative.

	{
		{
			\begin{figure}[ht!]
				\begin{subfigure}{.5\textwidth}
					\centering
					\hspace*{-1cm}\includegraphics[width=1.\linewidth]{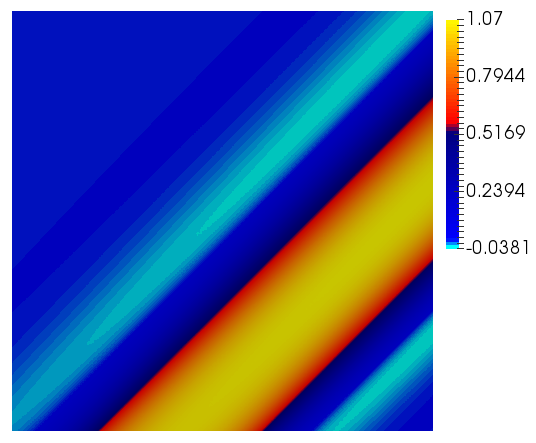}
					\caption{LS result in void.}
					\label{f:ls-void}
				\end{subfigure}
				~
				\begin{subfigure}{.5\textwidth}
					\centering
					\includegraphics[width=1.\linewidth]{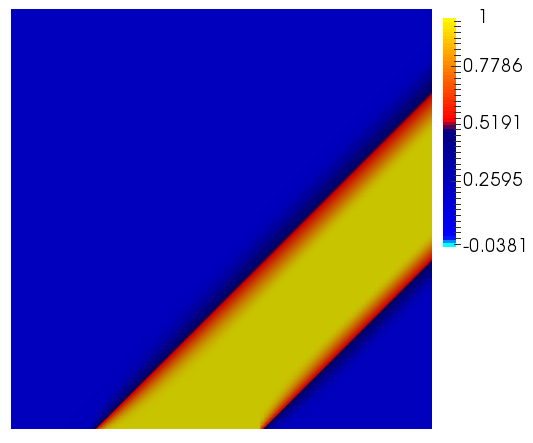}
					\caption{R\lo\ result in void.}
					\label{f:l1-void}
				\end{subfigure}
				\caption{LS and R\lo\ method comparison in void transport problem with a grazing, incident flux on the boundary. 100\tm100\ cells are used}
				\label{f:void-pcolor}
			\end{figure}
			
This problem can also demonstrate the necessity of having a boundary condition based on \lo. In Figure\ \ref{f:void-bd} we compare the analytic, LS, and R\lo\, solutions along the boundary at $y=0$. In the R\lo\, solution we use the \lo\ boundary condition we derived above, and the standard LS boundary condition. As seen in the figure, the LS solution does over and under shoot the analytic solution. When the R\lo\ method is used with the LS boundary condition, the result is a sharp oscillation near the discontinuity. These oscillations go away when the \lo-based boundary condition is used.
			
			\begin{figure}[ht!]
				\centering
				\hspace*{0cm}\includegraphics[width=.7\linewidth]{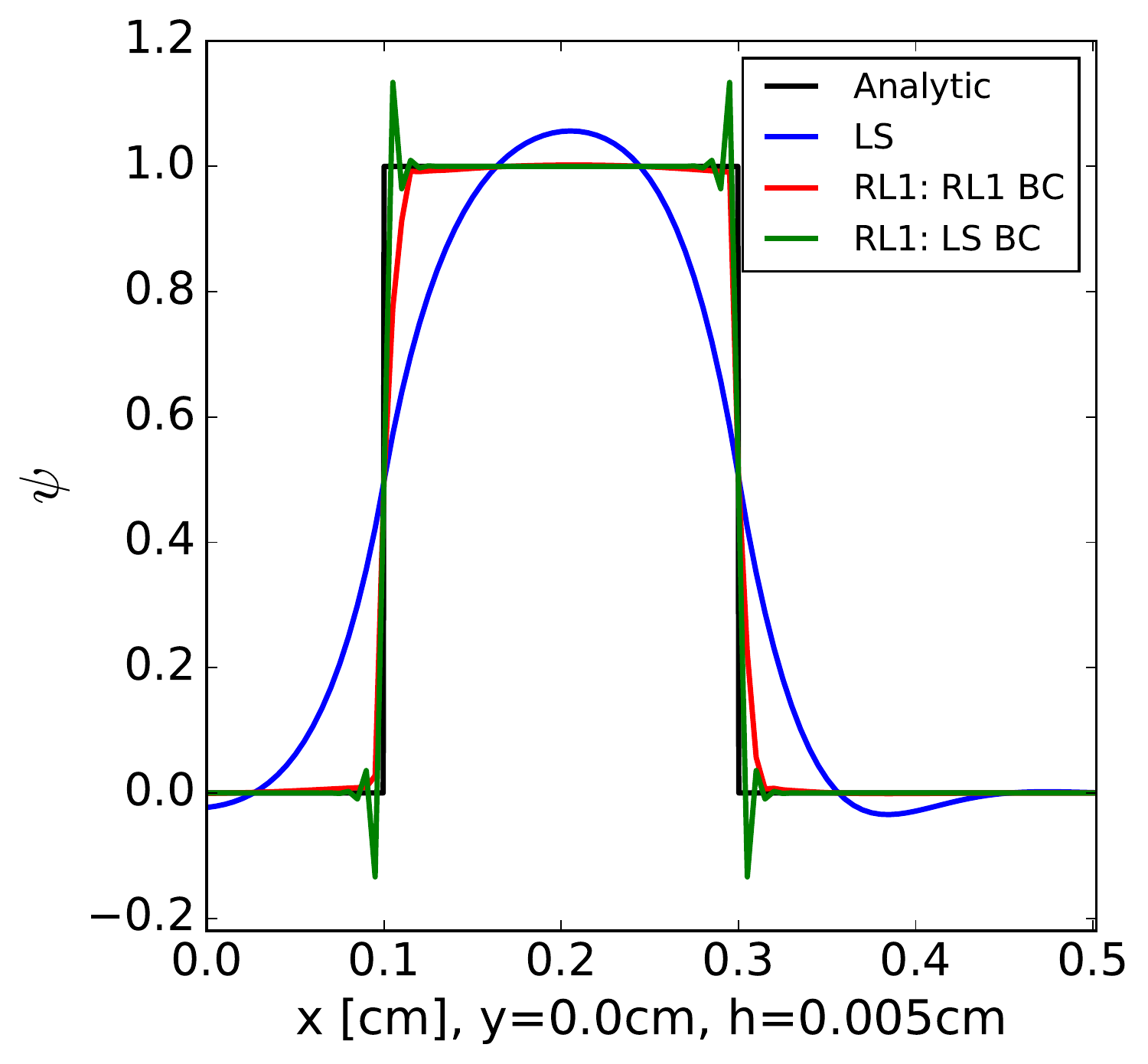}
				\caption{LS and R\lo\ with different boundary conditions on the void problem.}
				\label{f:void-bd}
			\end{figure}

The results in the previous two figures were performed with $\theta/|R|_\mathrm{max}=0.01$. In Figure\ \ref{f:void-theta} we show the line out of the solution $y=0.2$\ cm with differing values of $\theta/|R|_\mathrm{max}$. With a large value of $\theta$\ (the one with $\theta/|R|_\mathrm{max}=0.1$),\ R\lo\ is able to effectively damp the oscillations around the discontinuous solution, though the solution does become slightly negative. The results with $\theta/|R|_\mathrm{max}=0.01$ and $0.0001$ are nearly indistinguishable in the figure. For this reason we will use $\theta/|R|_\mathrm{max}=0.01$ for the remainder of this work.
			
			\begin{figure}[ht!]
				\centering
				\hspace*{0cm}\includegraphics[width=.7\linewidth]{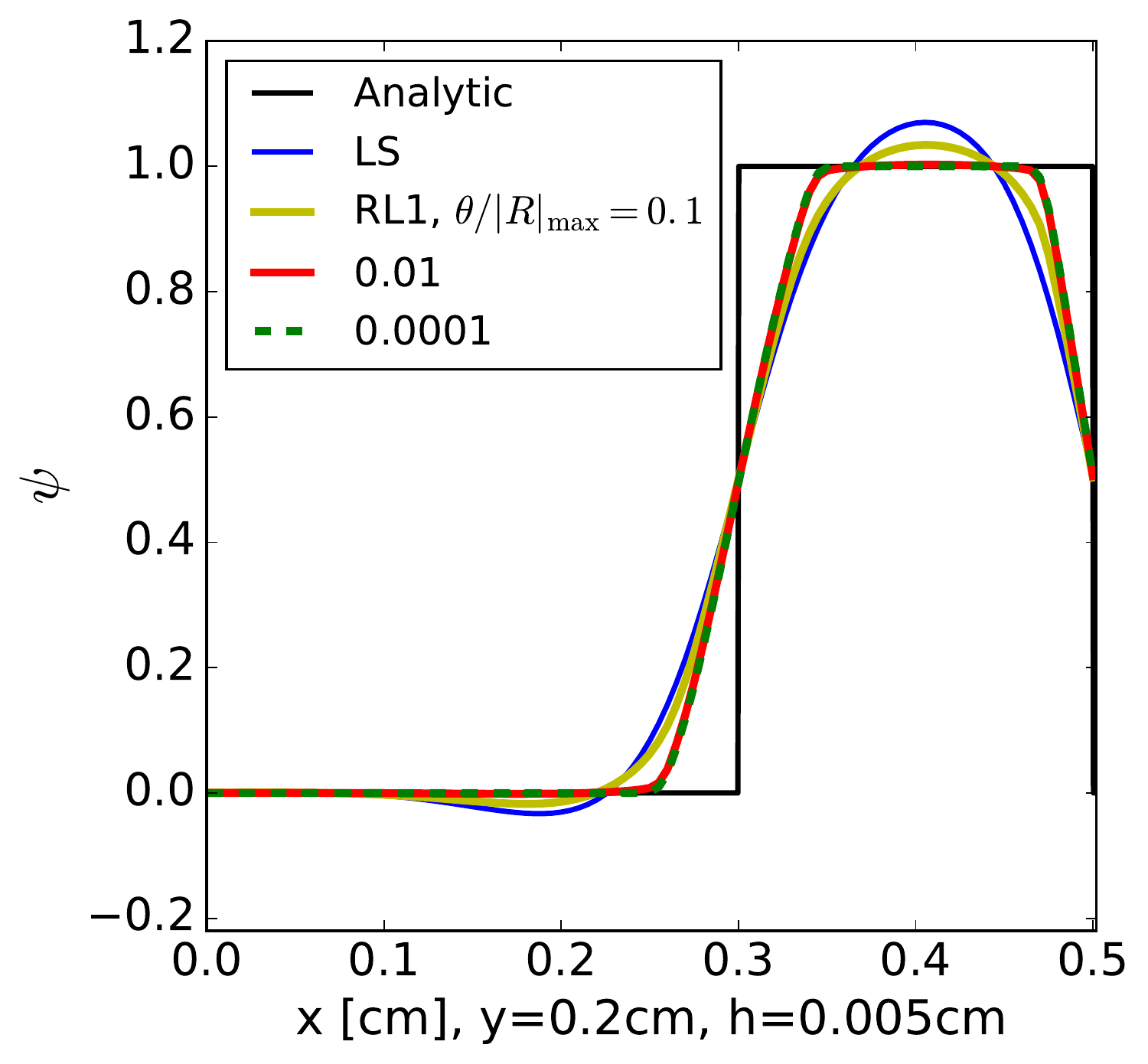}
				\caption{Solution to the void problem at  $y=0.2$\ cm for different values of $\theta/|R|_\mathrm{max}$.}
				\label{f:void-theta}
			\end{figure}

			As the mesh is uniformly refined we observe that the errors in the solution to the void problem decrease as the mesh size $h$ to the one-half power for both methods, as illustrated in Figure\ \ref{f:vd_conv},\ yet, the constant is smaller for the R\lo\ solutions. The converged R\lo\ solutions have the smallest error. Typically, a converged solution requires around $40$\ nonlinear iterations with $tol=1\e{-4}$\ used in this work. But only a few nonlinear iterations can provide a noticeable improvement over LS.
			
			\begin{figure}[ht!]
				\centering
				\hspace*{0cm}\includegraphics[width=.7\linewidth]{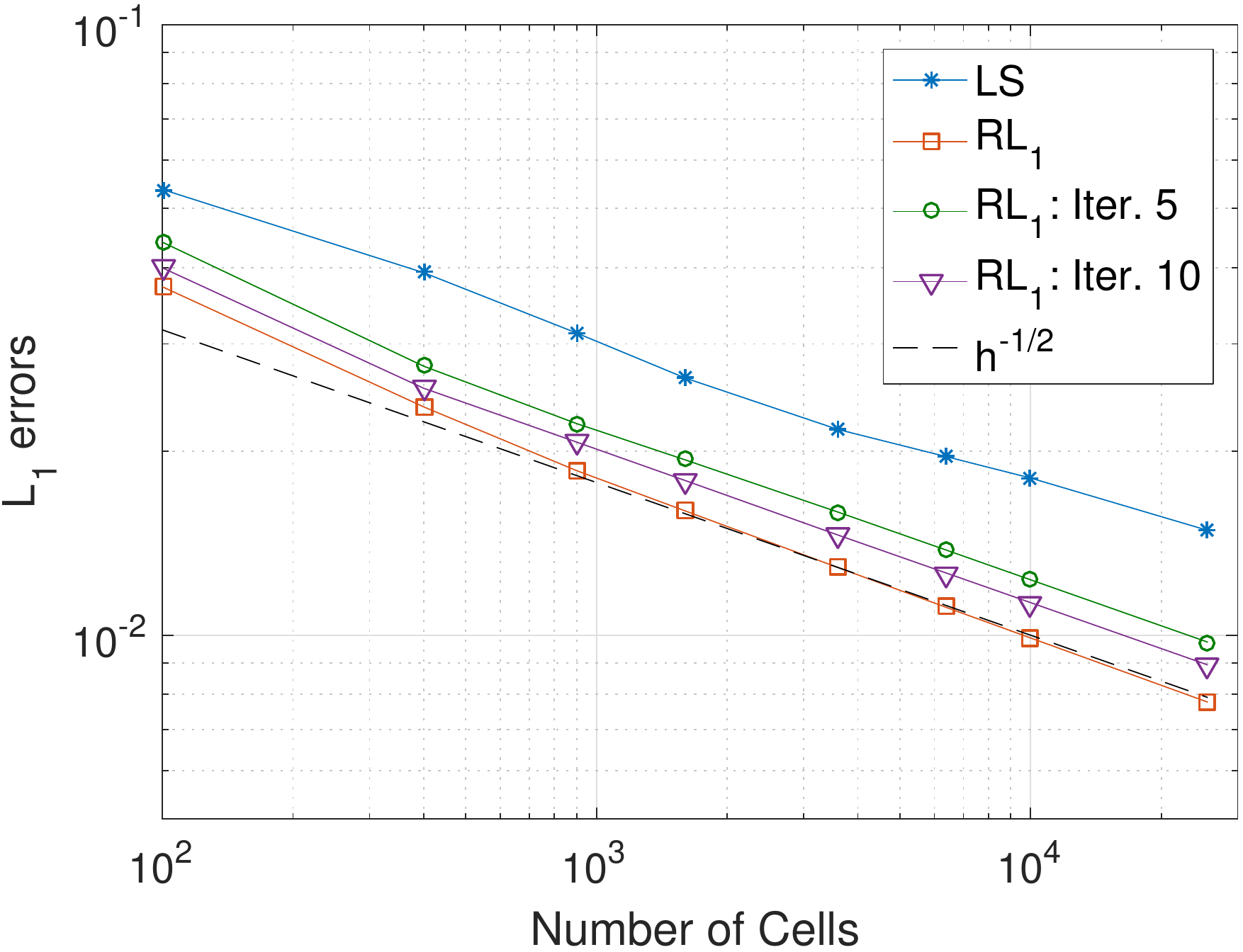}
				\caption{Void problem convergence results.}
				\label{f:vd_conv}
			\end{figure}
		}
		\subsection{Pure absorber problem}\label{s:abs}
		{
			The second test is with the same geometry as in the previous problem with the void replaced by an absorber with $\st=1.0$\ cm$^{-1}$.\ Additionally, the unit incident angular flux is imposed through the whole bottom boundary with the same incident angle.
						Figures\ \ref{f:abs-ls}\ and \ref{f:abs-l1}\ present the angular flux for this problem. In the pseudo-color plots in Figure \ref{f:abs-l1} we observe the oscillations and  negative values in the LS solution; these artifacts do not appear in the R\lo\ solution. Figure\ \ref{f:abs-line}\ compares R\lo\ and LS line-outs along y=0.2\ cm with the analytic solution. Indeed, neither of these methods resolves the discontinuity. However, the R\lo\ solution is monotonic, in contrast to LS's oscillatory flux as a result of Gibbs phenomenon.
We examine the \lo\ error of scalar fluxes as well. As the convergence tests in Figure\ \ref{f:abs-err}\ shows, we found both LS and R\lo\ have a convergence order near one half, with R\lo\ having lower error magnitudes as expected.
			
			\begin{figure}[ht!]
				\begin{subfigure}{.5\textwidth}
					\centering
					\hspace*{0cm}\includegraphics[width=1.\linewidth]{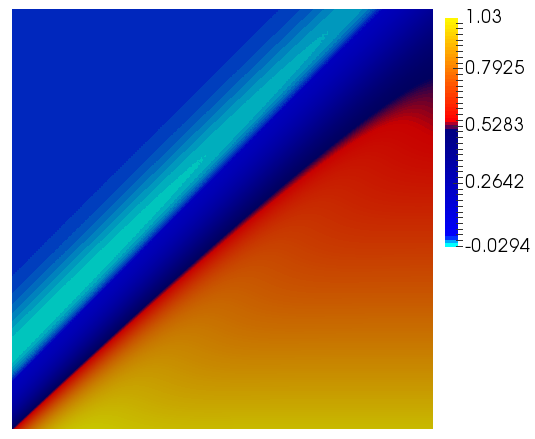}
					\caption{Least-Squares.}
					\label{f:abs-ls}
				\end{subfigure}
				~
				\begin{subfigure}{.5\textwidth}
					\centering
					\includegraphics[width=1.\linewidth]{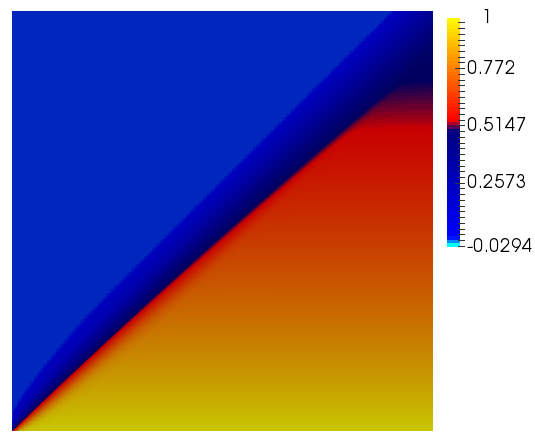}
					\caption{Regularized \lo.}
					\label{f:abs-l1}
				\end{subfigure}
				\caption{Angular flux distributions in incident convergence test.}
			\end{figure}
			
			\begin{figure}[ht!]
				\begin{subfigure}{.5\textwidth}
					\centering
					\hspace*{0cm}\includegraphics[width=1.\linewidth]{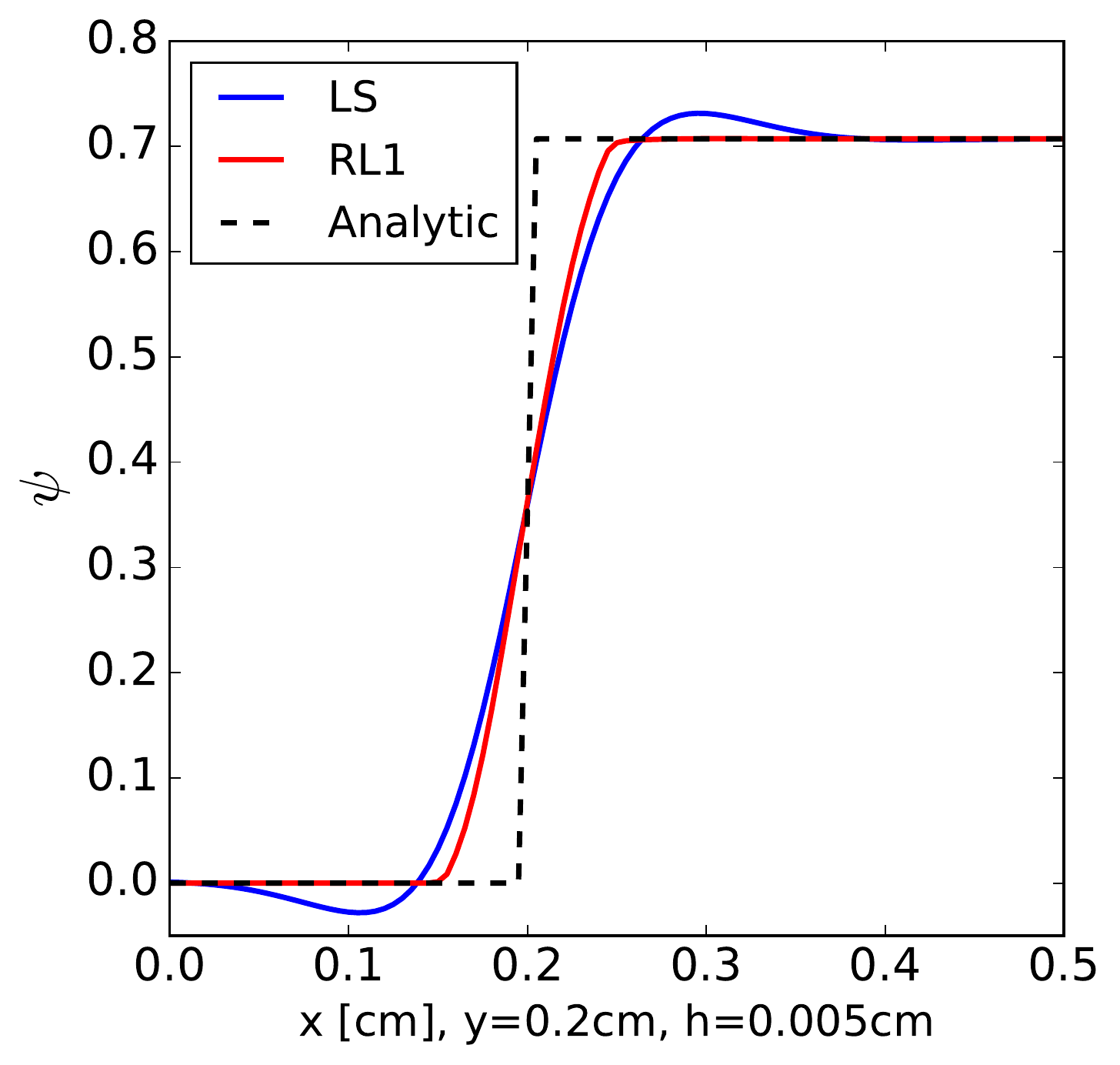}
					\caption{Angular flux line-outs along $y=0.2$\ cm.}
					\label{f:abs-err}
				\end{subfigure}
				~
				\begin{subfigure}{.5\textwidth}
					\centering
					\hspace*{0cm}\includegraphics[width=1.\linewidth]{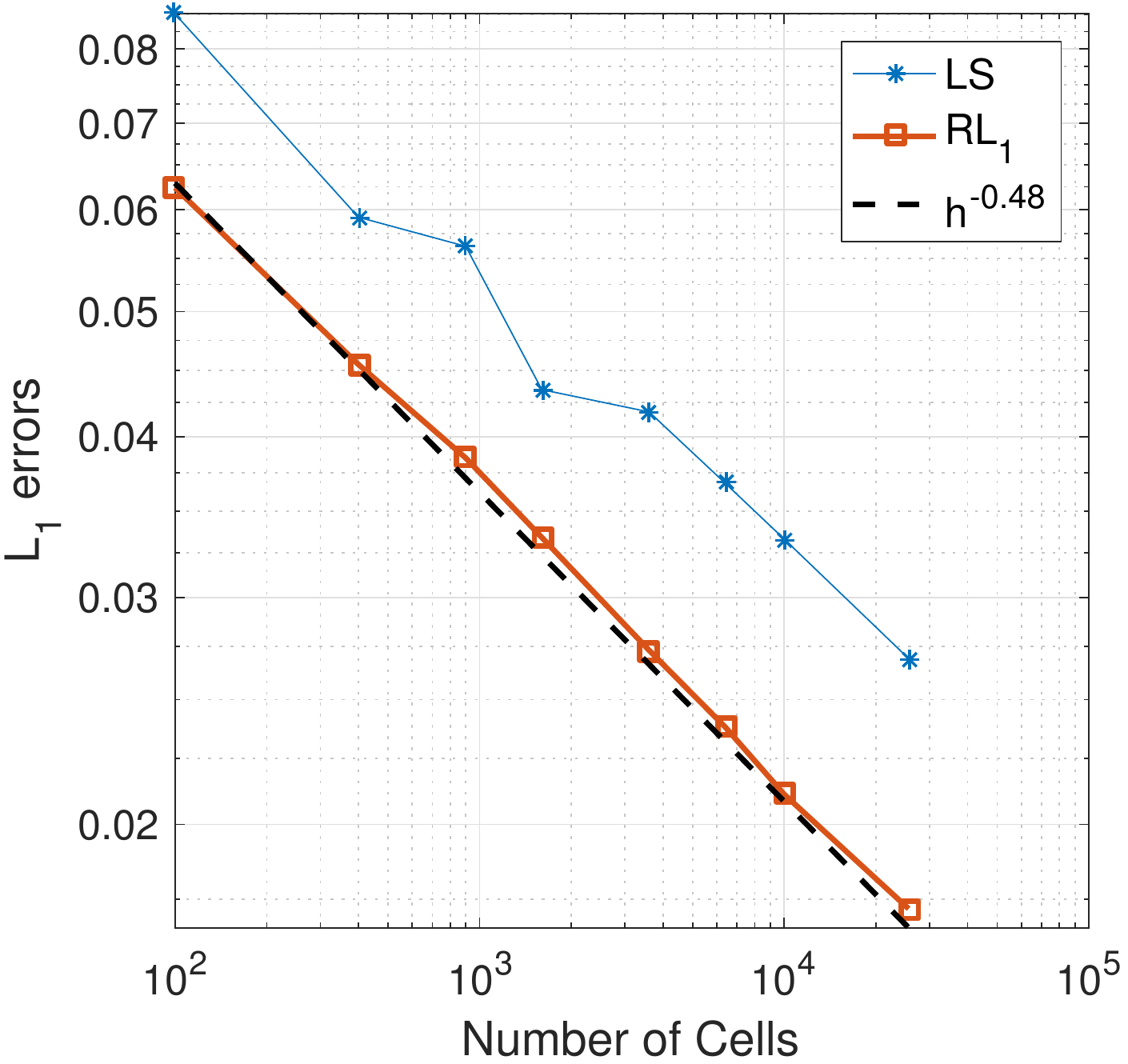}
					\caption{\lo\ norm of scalar flux errors.}
					\label{f:abs-line}
				\end{subfigure}
				\caption{Angular flux line-out and scalar flux errors for the pure absorber test problem.}
			\end{figure}

		}
	}
	
	\subsection{Smooth boundary problem}
	{
		In this test, we still use the same material and geometry configurations as in Sec.\ \ref{s:abs}\ except that the we use a smooth boundary condition specified at a grazing direction, $\Omega_x=0.8688,~\Omega_y=0.3599,\Omega_z=0.3599$. The boundary condition for this angle is
		\begin{align}
		\psiinc{}=\begin{cases}
		\displaystyle\frac{1}{2}+\frac{1}{2}\cos\left(2\pi\frac{x-0.2}{0.2}\right),&x\in(0.1,~0.3)\ \mathrm{cm},\ y=0\ \mathrm{cm},\\
		0, &\mathrm{otherwise}.
		\end{cases}
		\end{align}
		
		\begin{figure}[ht!]
			\centering
			\begin{subfigure}{.45\textwidth}
				\centering
				\hspace{-1cm}\includegraphics[width=1.\linewidth]{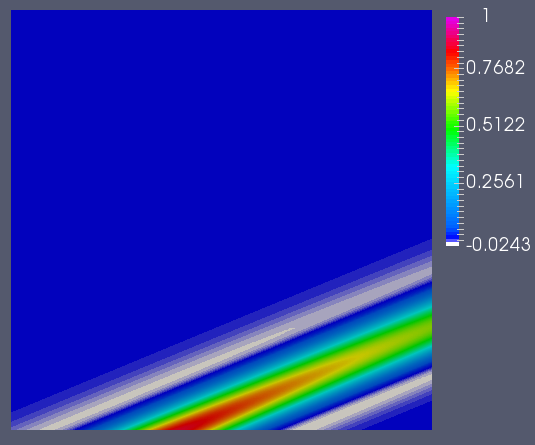}
				\caption{LS.}
				\label{f:sbc-ls}
			\end{subfigure}
			~
			\begin{subfigure}{.45\textwidth}
				\centering
				\hspace{1cm}\includegraphics[width=1.\linewidth]{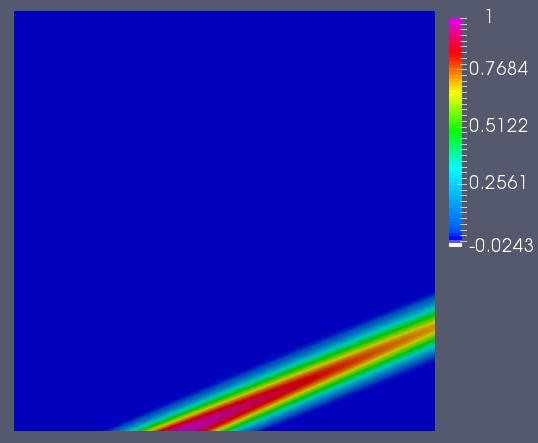}
				\caption{R\lo.}
				\label{f:sbc-l1}
			\end{subfigure}
			\caption{Angular flux distributions for the grazing incident.}\label{fig:smooth}
		\end{figure}
		
		Figure\ \ref{fig:smooth}\ illustrates the spatial distribution of angular flux for both methods. LS gives a negative angular flux even when the incident boundary condition is smooth in space.	
		As the line-out plots presented in Figs.\ \ref{f:abs-in}\ (incident boundary) and \ref{f:abs-out}\ (outgoing boundary),\ with converged R\lo,\ we see agreement with the analytic solution  except for slight smearing on the outgoing boundary in Figure\ \ref{f:abs-out}.\ Moreover, without convergence but with only one or two iterations of residual weighting, R\lo\ can still deliver acceptable results. On the other hand, LS solution presents negative solutions. Moreover, R\lo,\ in contrast to IRLS\cite{bnjiang_l1,lowrie_l1}, treats the smooth incident data correctly and propagates the information correctly throughout the domain.
		
		\begin{figure}[ht!]
			\begin{subfigure}{.5\textwidth}
				\centering
				\hspace*{-1cm}\includegraphics[width=1.\linewidth]{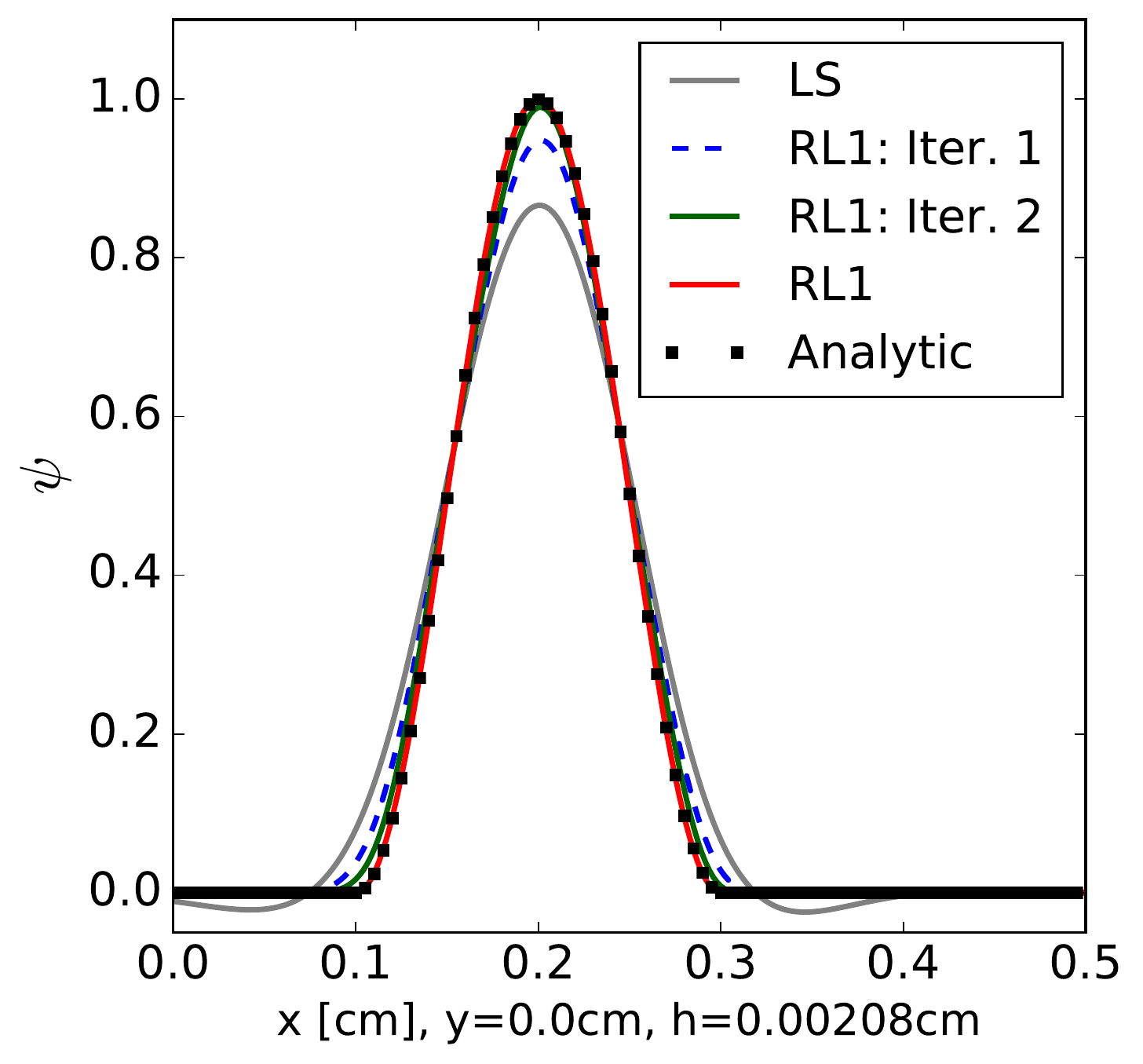}
				\caption{Incoming boundary (bottom).}
				\label{f:abs-in}
			\end{subfigure}
			~
			\begin{subfigure}{.5\textwidth}
				\centering
				\includegraphics[width=1.\linewidth]{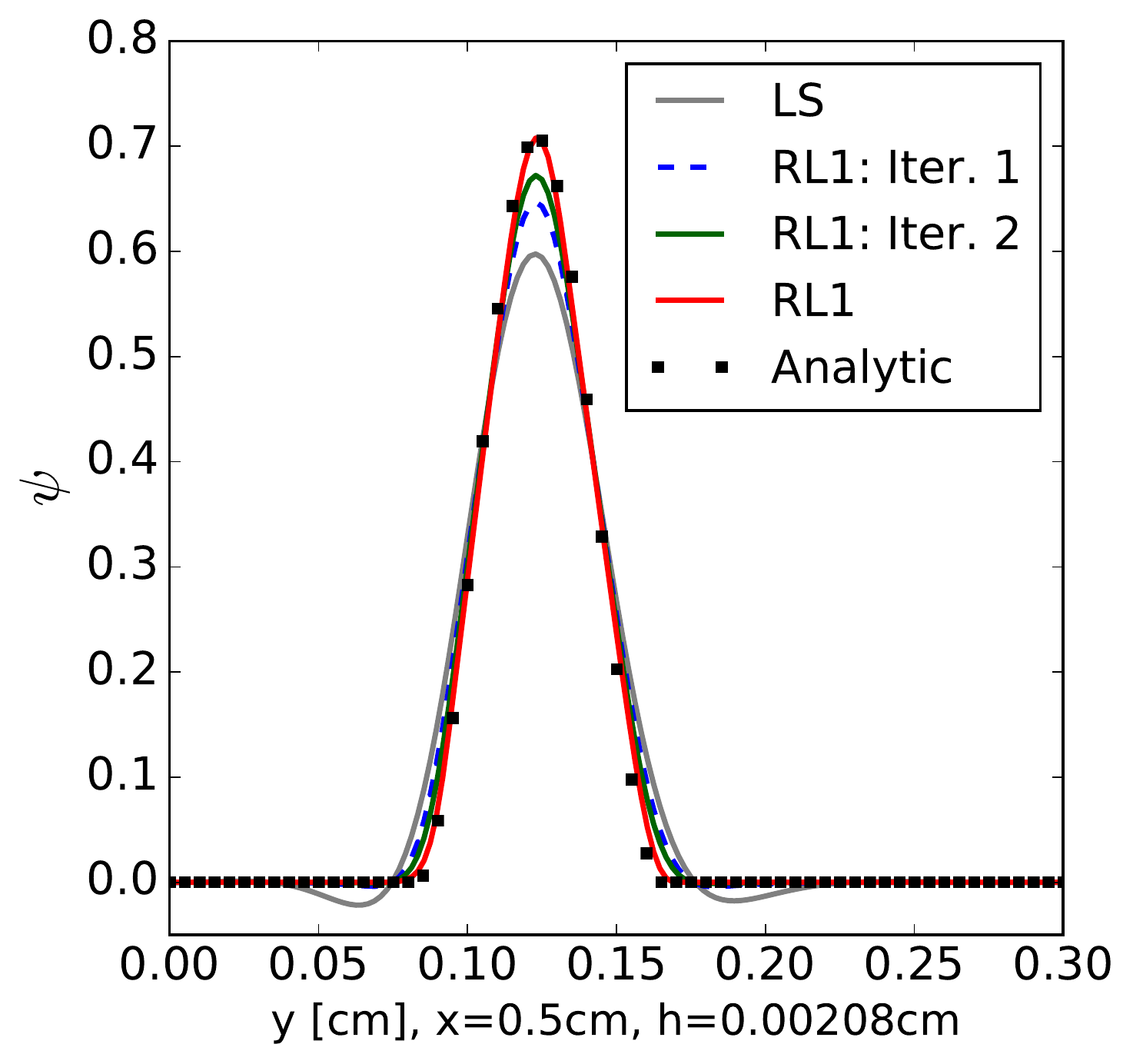}
				\caption{Outgoing boundary (right).}
				\label{f:abs-out}
			\end{subfigure}
			\caption{Boundary results comparison for smooth boundary problem.}
			\label{f:abs-comp}
		\end{figure}
	}

	\subsection{Ackroyd test}
	{
		Finally, we present the results for the Ackroyd problem\cite{ackroyd_void,clifmc}. This is a heterogeneous problem that has a high-scattering ratio central block and outer shield  ($\st=0.2,\sigs=0.19$). There is  a void between the block and the shield. See Fig.\ \ref{f:ack-config}\ for a schematic of a quarter of the problem geometry. The whole problem is symmetric about x-axis and y-axis. A reference scalar flux from an LS calculation using 320\tm320\ cells with a Gauss-Chebyshev level-symmetric-like \sss{8}\ quadrature\cite{jarrellmc}\ is presented in Figure\ \ref{f:ack-flux}.\ We examine the solution along a line that crosses the  void  in Figure\ \ref{f:ack-duct} and another one on the boundary in Figure\ \ref{f:ack-bd}.\ In both cases, the coarse mesh (32\tm32) R\lo\ solution agrees reasonably well with the fine mesh LS results, while the LS  coarse-mesh  solution has large and noticeable errors. When refining to 128\tm128,\ R\lo\ agrees with the reference, while LS still has large errors.
		
		\begin{figure}[ht!]
			\centering
			\begin{subfigure}{.4\textwidth}
				\hspace*{0cm}\includegraphics[width=1.\linewidth]{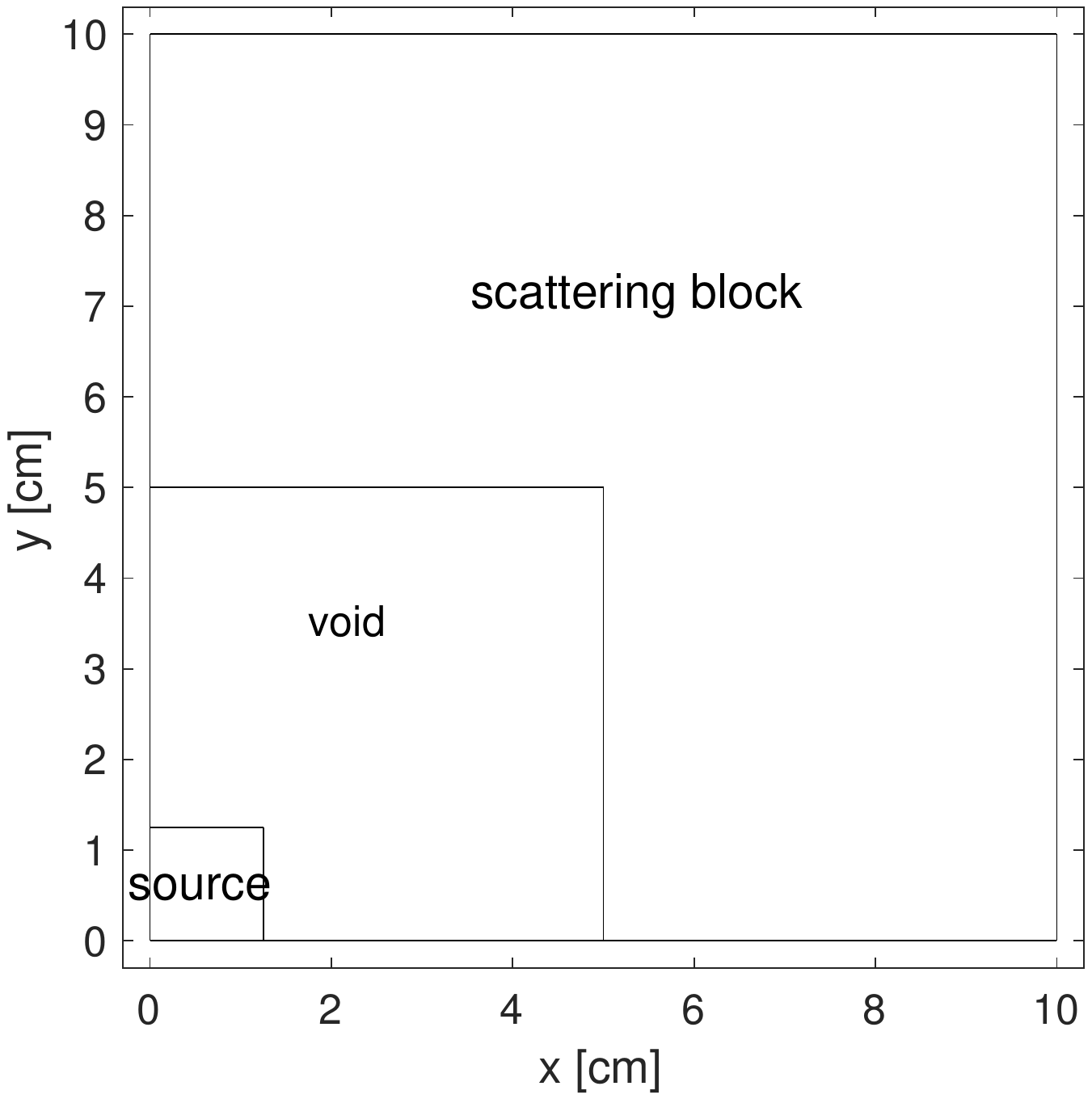}
				\caption{Ackroyd problem configuration.}
				\label{f:ack-config}
			\end{subfigure}
			~
			\begin{subfigure}{.5\textwidth}
				\hspace*{0cm}\includegraphics[width=1.\linewidth]{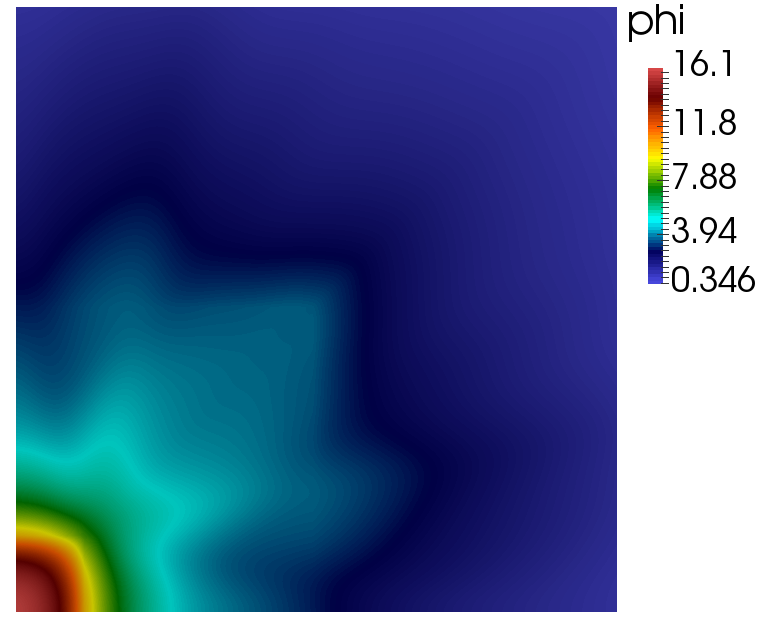}
				\caption{Ackroyd problem scalar flux distribution from LS solution with $320\times 320 $ cells and S$_8$\ quadrature.}
				\label{f:ack-flux}
			\end{subfigure}
			\caption{Ackroyd problem layout and reference solution.}
		\end{figure}
		
		\begin{figure}[ht!]
			\begin{subfigure}{.6\textwidth}
				\centering
				\hspace*{-0cm}\includegraphics[width=1.\linewidth]{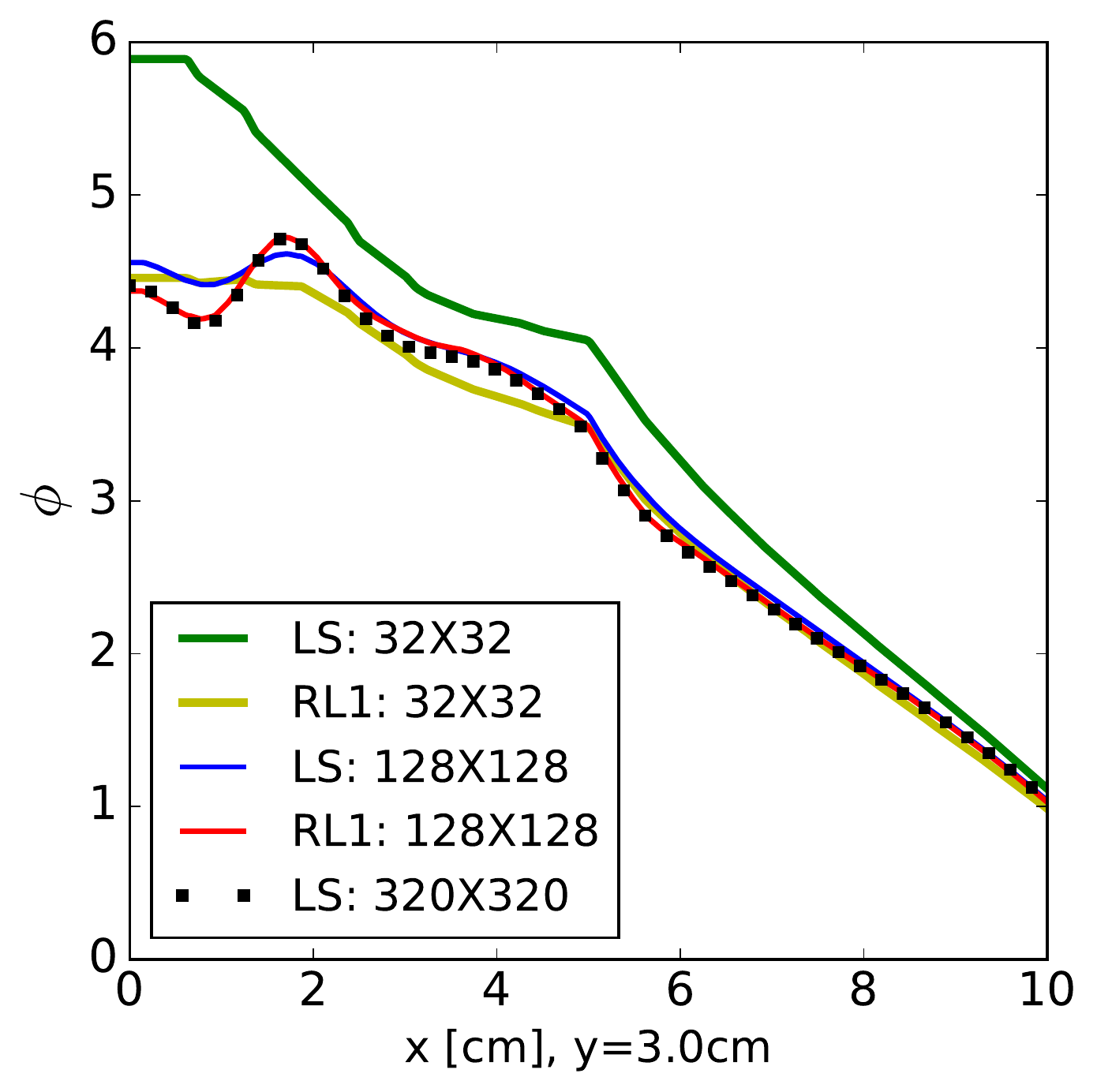}
				\caption{Solution across the void.}
				\label{f:ack-duct}
			\end{subfigure}
			~
			\begin{subfigure}{.6\textwidth}
				\centering
				\hspace*{-0cm}\includegraphics[width=1.\linewidth]{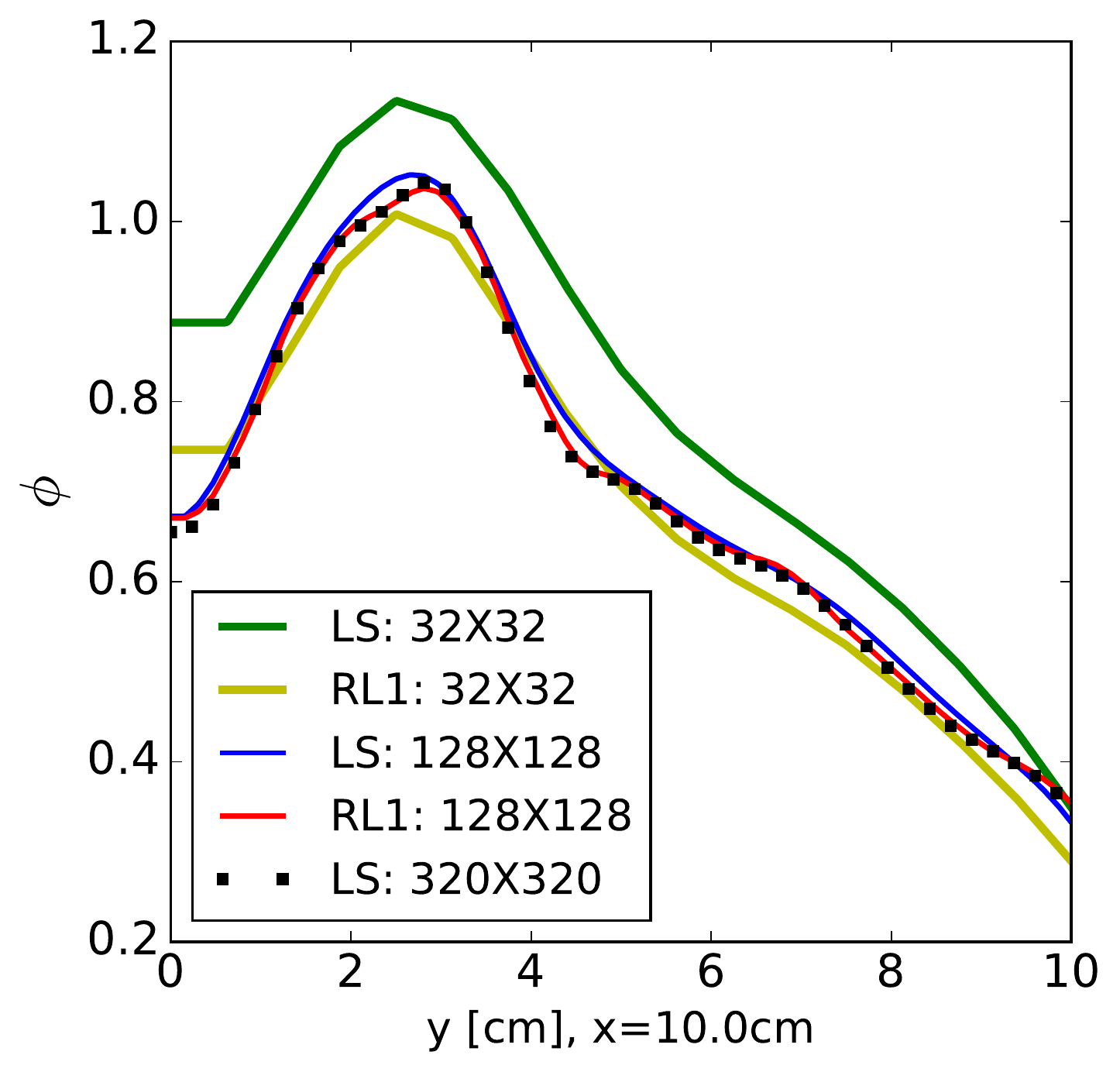}
				\caption{Solution on right boundary.}
				\label{f:ack-bd}
			\end{subfigure}
			\caption{Ackroyd problem line-out plots.}
			\label{f:ack}
		\end{figure}
	}
	
}

\section{Concluding remarks and future work}\label{s:conc}
In this work, we developed an effectively non-oscillatory regularized \lo\ finite element scheme for solving radiation transport problems. Starting from minimizing the \lo\ norm of transport residual, we derived a continuum form of \lo\ finite element method along with a consistent nonlinear \lo\ BC. The resulting continuum form is a weighted LS problem where the weight is the inverse of the absolute value of the residual. We regularize this form by switching the weight to unity when the residual is small. Numerical tests with void, pure absorber and heterogeneous problems demonstrate the efficacy and accuracy of our methodology.

There exist opportunities to improve our  method. For one, our technique for solving the resulting nonlinear equations is based on fixed-point iteration. The universal efficiency of such a scheme is to be investigated. Other possibilities, such as Jacobian-free Newton Krylov method\cite{knoll_jfnk}, could improve the efficiency of our scheme. Furthermore, our scheme inherits the properties of LS in that it is only conservative in the limit of the mesh size going to zero. Other authors have presented ways to ameliorate this issue (such as SAAF weighting for problems without void) and other void treatments \cite{laboure2016globally, yaqi_void}. Another possibility is to use a high-order low-order formalism \cite{bolding2017high,wieselquist2014cell} to solve the transport equation, where R\lo\ is used for the high-order solve.


\section*{Acknowledgements}
W. Zheng is thankful to Dr. Wolfgang Bangerth from Colorado State University for suggestions on the methodology derivation. Also, appreciation goes to Dr. Milan Hanus from Texas A\&M University for fruitful discussions and helpful proofreading.

This project is funded by Department of Energy NEUP research grant from Battelle Energy Alliance, LLC- Idaho National Laboratory, Contract No: C12-00281.
\section*{References}
\bibliography{mybibfile}

\end{document}